\documentclass[12pt,a4paper,superscriptaddress,preprint]{revtex4}

\usepackage[dvips]{graphicx}
\usepackage{amssymb,amsmath}
\usepackage{color}
\usepackage{bm}

\usepackage{enumerate}

\usepackage{booktabs}

\begin{document}

\newcommand{\EQ}{Eq.~}
\newcommand{\EQS}{Eqs.~}
\newcommand{\FIG}{Fig.~}
\newcommand{\FIGS}{Figs.~}
\newcommand{\TAB}{Tab.~}
\newcommand{\TABS}{Tabs.~}
\newcommand{\SEC}{Sec.~}
\newcommand{\SECS}{Secs.~}
\newcommand{\APDX}{Appendix~}

\title{Rich-club network topology to minimize synchronization cost due to phase difference among frequency-synchronized oscillators}

\author{Takamitsu Watanabe \footnote{takawatanabe-tky@umin.ac.jp}}
\affiliation{Department of Physiology, School of Medicine, The University of Tokyo,
7-3-1 Hongo, Bunkyo-ku, Tokyo 113-8656, Japan}

\begin{abstract}
Functions of some networks, such as power grids and large-scale brain networks, rely on not only frequency synchronization, but also phase synchronization. Nevertheless, even after the oscillators reach to frequency-synchronized status, phase difference among oscillators often shows non-zero constant values. Such phase difference potentially results in inefficient transfer of power or information among oscillators, and avoid proper and efficient functioning of the network. 
In the present study, we newly define synchronization cost by the phase difference among the frequency-synchronized oscillators, and investigate the optimal network structure with the minimum synchronization cost through rewiring-based optimization. 
By using the Kuramoto model, we demonstrate that the cost is minimized in a network topology with rich-club organization, which comprises the densely-connected center nodes and peripheral nodes connecting with the center module.
We also show that the network topology is characterized by its bimodal degree distribution, which is quantified by Wolfson's polarization index. 
Furthermore, we provide analytical interpretation on why the rich-club network topology is related to the small amount of synchronization cost.
\end{abstract}


\maketitle

\section{Introduction}\label{Sec:Introduction}
As power grids \cite{Kundur1993Elgerd1982Book, Dorfler2010ACC} and networks of bursting neurons \cite{Varela2001NatRevNeuro, Fell2011NatRevNeuro}, functions of some complex networks of oscillators are based on not only synchronization of frequencies of the oscillators, but also synchronization of their phases. However, in general, frequency synchronization is more achievable than phase synchronization. Phase difference among frequency-synchronized oscillators often falls into a non-zero constant, and such non-zero phase difference would avoid proper and efficient functioning of the complex networks. 

In power grids, alternating voltage of the power plants in the grids should be synchronized around certain specific frequency (e.g., 50 Hz in the most parts of Europe and 60 Hz in the north America) \cite{Kundur1993Elgerd1982Book}, and disruption of the frequency synchronization causes a blackout in a large area \cite{Kundur1993Elgerd1982Book, HillChen2006IEE}. In addition, the phases of the voltages of the power plants are also required to be synchronized. As discussed in \APDX\ref{Appendix: SyncCostInPowerGrid}, the difference in voltage phases among power plants inevitably causes power loss consumed as heat in power lines \cite{Kundur1993Elgerd1982Book, Dorfler2010ACC}. 
In this sense, the phase difference in power grids can be regarded as synchronization cost.
Considering recent increasing environmental awareness and soaring global demand of natural resources \cite{USEFC2008, SmartGrid2011IEEE}, it is necessary to reduce the synchronization cost due to the phase difference among frequency-synchronized power plants.
%

Synchronization in large-scale brain networks also requires less difference in phase of neuronal activity among different brain regions. 
A series of prior experimental researches have shown that various important functions in large-scale brain networks are based on not only frequency-  but also phase- synchronization\cite{Varela2001NatRevNeuro, Fell2011NatRevNeuro}. 
A previous electrophysiological study showed that spike activity recorded from monkeys' cortex exhibited phase synchronization in various frequency bands while the monkeys were conducting tasks that required integration of visual processing and motor responses \cite{Bressler1993Nature}. 
In another careful electrophysiology study, Roelfsema and his colleagues recorded local field potentials (LFP) in the cerebral cortex in cats, and revealed that phase synchronization between LFPs recorded in the visual and parietal cortices was increased only when the cats focused their attention on stimuli \cite{Roelfsema1997Nature}. 
Studies using electro-encephalogram to record human brain activity found that increase of phase synchronization in various frequency was associated with learning and perception of images \cite{Miltner1999Nature, Hipp2011Neuron}.
Furthermore, the frequency- and phase- synchronization are considered to occur in large-scale brain networks \cite{Bressler1993Nature, Traub1996Nature} with zero time-lag \cite{Roelfsema1997Nature, Womelsdorf2007Science, Fell2011NatRevNeuro}. 
These researches suggest that phase synchronization in large-scale brain networks is related to crucial functions such as integration of multiple information \cite{Varela2001NatRevNeuro}, neural communication \cite{Fries2005TrendsCogSci}, and spike-timing-dependent plasticity \cite{Fell2011NatRevNeuro}. Actually, it is known that some types of the disruption of the synchronization cause dysfunctions of memory learning \cite{Stopfer1997Nature} or psychiatric disorders \cite{Tononi2000RBRR}.
Considering these findings, it is to some extent reasonable to hypothesize that large-scale brain networks have a specific organization that minimizes phase differences among brain activity, and enables optimal phase synchronization in the entire networks.
%

These previous literatures indicate the importance to reduce phase difference among frequency-synchronized oscillators, which can be regarded as a type of cost that is spent during synchronization. However, little is examined about the optimal network topology that reduces this type of synchronization cost due to phase difference. 
Indeed, a series of previous literatures have investigated optimal network structures by introducing a different type of synchronization cost, which is needed for building or maintain of the optimal network infrastructure. A study regarded parameters based on coupling strength among oscillators as a cost, and demonstrated that homogeneous and uniform distribution of the coupling strength enhances the tendency of synchronization \cite{Motter2005PRE}. Another study employing the same definition of synchronization cost suggests that more heterogeneous network structures are required to simultaneously achieve both the maximum synchronizability and minimum synchronization cost based on coupling strength \cite{Nishikawa2006PhyD}. Another study revealed the optimal distribution of coupling strength that increases synchronizability among Kuramoto oscillators \cite{Brede2008EPJB}. 
However, these prior researches have not focused on synchronization cost due to phase difference. 
Consequently, despite a line of prior researches on optimal conditions for synchrony in networks \cite{Arenas2008RhysRep}, it remains unclear about optimal network structures that minimize synchronization cost due to phase difference among oscillators.

In the present study, therefore, we examine the optimal network topology to minimize the phase-difference-based synchronization cost. 
We take the following five steps:
\begin{enumerate}[(i)]
\item First, we define the synchronization cost, $S_{ij}$, due to phase difference between frequency-synchronized phase oscillators $i$ and $j$ in the Kuramoto model. We adopt the Kuramoto model because the model has been used as approximation of various systems including power grids \cite{Buzna2009PRE, Fioriti2009CIIS, Dorfler2010ACC} and large-scale brain networks \cite{Breakspear2010FrontHN, GomezGardenes2010PLosOne}. 
\item Second, by using the definition, we numerically calculate the mean of synchronization cost, $\left< S \right>$, in the entire network.
\item Third, by using a rewiring strategy \cite{Donetti05PRL}, we show that "rich-club" network topology \cite{Zhou2004IEEEComm, Colizza2006NatPhys, GomezGardenes2010PLosOne, Heuvel2011JNS}, which consists of densely inter-connected modules and peripheral low-degree nodes, is the optimal topology with the minimum synchronization cost.
\item Forth, we characterize the rich-club network topology by quantifying the bimodality of the degree distribution of the network.
\item Finally, we provide an analytical interpretation on why the rich-club network organization is associated with a small amount of the synchronization cost.
\end{enumerate}

\section{Method}

\subsection{Definition of synchronization cost}
We first define synchronization cost due to phase difference in an unweighted and undirected network which is described by an adjacency matrix $A$ and consists of $N$ phase oscillators. $A_{ij}$ is $1$ when oscillators $i$ and $j$ are connected, and $A_{ij}$ is $0$ when they are not. 
According to the Kuramoto model, the phase of the oscillator $i$, $\theta_i$, is described as
\begin{equation}
	 \dot{\theta_i} = {\omega_i} - \epsilon \sum_{j}{A_{ij} \sin(\theta_i - \theta_j)},
	 \label{eq:KuramotoEq}
\end{equation}
where $\omega_i$ is the natural frequency of the oscillator $i$, and $\epsilon$ is a coupling strength. To reduce computational cost for the following rewiring-based optimization, we assume that $\epsilon$ is constant for any combination of oscillators. 
In this Kuramoto model, the synchronization cost between oscillators $i$ and $j$, $S_{ij}$ is defined based on the phase difference between the connected oscillators as follows:
\begin{equation}
	S_{ij} = \left( \theta_i - \theta_j \right)^2.
\end{equation}
Note that the synchronization cost is only defined after the network of the oscillators reaches to a state of frequency synchronization. As described in \APDX\ref{Appendix: SyncCostInPowerGrid}, in power grids, $S_{ij}$ can be regarded as an index that is proportional to power loss due to difference in voltage phase between power plants $i$ and $j$.

\subsection{Estimation of the mean synchronization cost}
Based on the definition of $S_{ij}$, we numerically estimate $S_{ij}$ for each edge in the following four steps for a given network:
\begin{enumerate}[(i)]
\item We set normally-distributed $\{ \omega_i \}$ for each node. It is because that previous studies on real networks such as power grids and brain networks have assumed that the natural frequencies of belonging oscillators are symmetrically fluctuating around the averaged frequency \cite{Filatrella2008EPJB, Fioriti2009CIIS, Fell2011NatRevNeuro}.
\item Based on the Kuramoto model described in Eq.~\eqref{eq:KuramotoEq}, we numerically estimate frequency-synchronized status, where $\dot{\theta_i}$ becomes a common constant value, $\Omega$, for any $i$. 
\item In the frequency-synchronized status, each oscillator has a different specific phase, $\theta_i$. Based on the set of $\{ \theta_i \}$, we then evaluate $S_{ij}$ for each edge. 
\item As described in (i), the set of the natural frequency, $\{ \omega_i \}$, is fluctuating over time. Thus, the $S_{ij}$ is also fluctuating over time. Therefore, we repeat the procedure (i)-(iii) 200 times with different sets of $\{ \omega_i \}$, and obtain 200 different sets of $\{ S_{ij} \}$. Then, we average the $\{ S_{ij} \}$ over time, obtaining $\left< S_{ij} \right>$ for each edge. Finally, we average the $\left< S_{ij} \right>$ across edges, and obtain $\left< S \right>$ for the entire network.

\end{enumerate}

\subsection{Rewiring-based optimization}
To search for the optimal network topology with the least $\left< S \right>$, we adopt the rewiring method that previous studies used to explore the network topology with the largest synchronizability \cite{Donetti05PRL}.
We apply the following rewiring-based optimization procedure to a given connected network with $N$ nodes and mean degree of $\left< k\right>$: At each step, the number of rewired edges is randomly determined based on an exponential distribution. 
The set of edges to be rewired is also randomly chosen in a given network. 
After the rewiring, we estimate frequency-synchronized status and obtain $\left< S \right>_{\rm updated}$. 
The attempted rewiring is rejected if the updated network is disconnected. 
Otherwise, the rewiring is accepted 
if $\Delta \left< S \right> = \left< S \right>_{\rm updated} - \left< S \right>_{\rm initial} < 0$, 
or with probability $p = \min (1, [1-(1-q) \Delta \left< S \right> / T]^{1/(1-q) } )$ where $T$ is a temperature-like parameter and $ q = -3 $ \cite{Donetti05PRL}. 
The initial rewiring is conducted at $T = \infty $, and, after the first $N$ rewiring, $T$ is set as $(1-q) (\Delta \left< S \right>)_{\max}$ where $(\Delta \left< S \right>)_{\max}$ is the largest $ \Delta \left< S \right> $ in the first $N$ rewiring trials. After that, $T$ is decreased $10\%$ in every $10$ rewiring trials. This estimation process iterated until there is no change in more than 50 successive rewiring steps.
We apply this rewiring-based optimization to three different initial networks: an Erd\H{o}s-R\'{e}nyi (ER) random model, Watts-Strogatz (WS) model \cite{Watts98Nature}, and a Barab\'{a}si-Albert (BA) model \cite{Barabasi99Science} with $N = 50$ and $\left<k\right> = 4$ \cite{Donetti05PRL}. 
In all the cases, the coupling strength, $\epsilon$, is set to be $0.3$. Each set of the natural frequencies of Kuramoto oscillators, $\{ \omega_i \}$, is randomly chosen from the normal distribution with an average of $100 \pi$ and a standard deviation of $1$.

During the optimization, we trace the standard order parameter, $r$, and local synchoronizability, $r_{\rm local}$ \cite{Gomez-Gardenes2007PRLPRE}, defined as follows:
\begin{gather}
	r e^{i \psi} = \frac{1}{N} \sum_{j}{e^{ i \theta_j}}, \label{eq:OrderEq} \\
	r_{\rm local} = \frac{1}{2N_l} \sum_{i} \sum_{j \in \Gamma_i} 
		\Biggl| \lim_{\Delta t \to \infty} \frac{1}{\Delta t} \int_{t_r}^{t_r + \Delta t} 
		e^{i [\theta_i (t) - \theta_j (t) ]} dt  \ \Biggr|,
\end{gather}
where $N_l$ is the total number of edges, $\Gamma_i$ is the set of neighbors of node $i$, and $\psi$ is a global phase.
Furthermore, after the optimization is completed, we compare the optimized networks from the different initial networks by estimating the following basic topological properties: mean of shortest path length \cite{Bunde1991Book}, mean of clustering coefficient \cite{Watts98Nature}, mean of betweenness centrality \cite{Newman2001PRE}, and degree correlation \cite{Newman2001PRE}. We conducted ten optimizations of ten different networks for each type of the initial networks, and averaged these basic properties.
\begin{itemize}
\item The shortest path length, $\ell_{ij}$, is defined as the shortest distance between two nodes $i$ and $j$ \cite{Bunde1991Book}. The averaged shortest path length, $\left< \ell \right>$, is defined as the average value of $\ell_{ij}$ over all the possible pairs of nodes in the network.
\item The clustering coefficient for node $i$, $C_i$, measures the local group cohesiveness \cite{Watts98Nature}, which is defined as the ratio of the number of links between the neighbors of $i$ and the maximum number of such links. We define the mean clustering coefficient, $\left< C \right>$, as an average value over all the nodes.
\item The betweenness centrality for node $i$, $b_i$, is defined as the number of shortest paths between pairs of nodes that pass through a given node \cite{Newman2001PRE}. We define the mean betweenness centrality, $\left< b \right>$, as an average value over all the nodes.
\item The degree correlation for a network is defined as the Pearson assortativity coefficient of the degrees, $r_{\rm assortative}$ \cite{Newman2001PRE}. The coefficient enables us to quantify the preference for high-degree nodes to attach to other high-degree nodes. Networks with this preference show large $r_{\rm assortative}$. 
\end{itemize}

\subsection{Estimation of rich-club coefficient}
We estimate reich-club coefficient, $\Phi (k)$, for both initial networks and optimized networks. According to the previous studies \cite{Zhou2004IEEEComm, Colizza2006NatPhys, GomezGardenes2010PLosOne, Heuvel2011JNS}, the coefficient for each degree $k$ is calculated as
\begin{equation}
	\Phi (k)  = \frac{2 E_{>k}}{N_{>k} (N_{>k} -1)},
	\label{eq:RichClubCoeff}
\end{equation}
where $E_{>k}$ represents the number of edges among $N_{>k}$ nodes that have more than $k$ degrees. As in previous studies \cite{Colizza2006NatPhys, GomezGardenes2010PLosOne, Heuvel2011JNS}, we calculate normalized rich-club coefficients, $\Phi_{\rm norm} (k)$ through dividing the raw value, $\Phi (k)$, by the mean of rich-club coefficients of 100 random networks (ER models), $\Phi_{\rm random} (k)$, as follows,
\begin{equation}
	\Phi_{\rm norm} (k)  = \frac{\Phi (k)}{\Phi_{\rm random} (k)}.
	\label{eq:RichClubCoeffNormalize}
\end{equation}
%

\section{Results}
\subsection{Rewiring-based optimization}\label{sec:Optimization}
\begin{figure}
\begin{center}
\includegraphics[width=15cm]{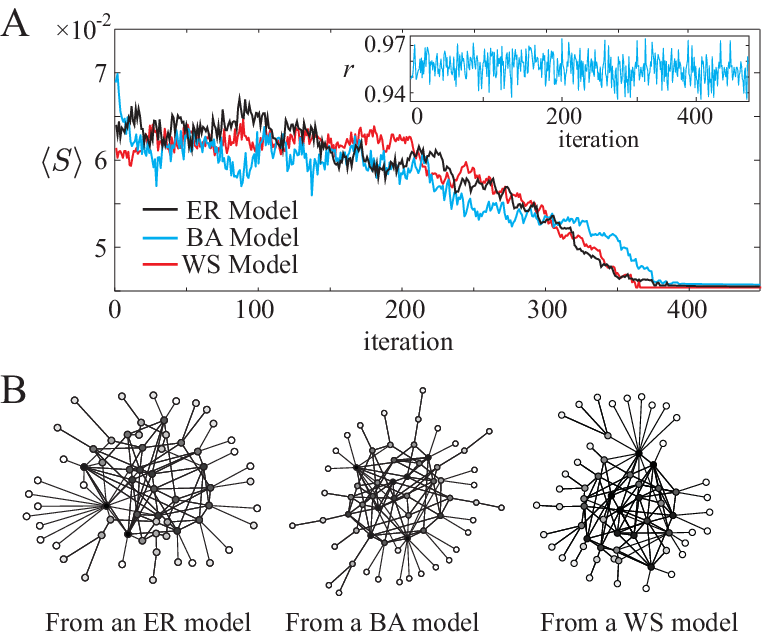}
\caption{
(Color online) 
{\bf A}. 
Main panel: Change of the synchronization cost, $\left< S \right>$, during rewiring-based optimization. Despite different initial networks (ER, BA, and WS models) with $N = 50, \left<k\right> = 4$, the synchronization cost converge to a similar amount of $\left< S \right>$.
Sub panel: Change of the standard order parameter, $r$, during the optimization. In contrast to $\left<S\right>$, the standard order parameter does not show notable change, just fluctuating below 1. The line shows the change of $r$ when the initial network is the BA model. In cases of the other two initial networks, the similar fluctuations were observed.
{\bf B}. Network topology optimized from different initial networks. Optimized networks are similar to each other. 
They have rich-club network topology, which consists of a densely-connected core module and peripheral low degree nodes connecting with the core. The color in the nodes represent the degree of the nodes: darker nodes have more edges.
}
\label{FigOptimizationResults}
\end{center}
\end{figure}

\FIG\ref{FigOptimizationResults} shows representative results of the rewiring-based optimizations. $\left< S \right>$ was decreased from approximately $6.5 \times 10^{-2}$ to $4.5 \times 10^{-2}$ even when the initial network structure was different. 
In all the initial networks, $\left< S \right>$ reached to a stable status after approximate 400 steps of rewiring. Strikingly speaking, we cannot guarantee that the optimal network was found, but this result suggests that a reasonably robust approximation of the optimal topology was obtained in this method.
%

During the optimization, the standard order parameter, $r$, were fluctuating just below 1 during the optimization (a small panel in \FIG\ref{FigOptimizationResults} A). The local synchonizability, $r_{\rm local}$, showed the similar fluctuation below 1. Considering the previous studies on these parameters \cite{Gomez-Gardenes2007PRLPRE}, these behaviors of $r$ and $r_{\rm local}$ are considered to be related to the amount of the coupling strength, $\epsilon$. 
The previous studies \cite{Gomez-Gardenes2007PRLPRE} have demonstrated that, when the coupling strength is more than 0.2, both of $r$ and $r_{\rm local}$ reach a plateau that is near to 1 regardless of network topology.
In the present study, the coupling strength, $\epsilon$, was set at 0.3, because global synchronization is necessary for the estimation of $\left< S \right>$. This relatively large coupling strength could result in the saturation of the global and local order parameters, $r$ and $r_{\rm local}$ during the optimization process.

\FIG\ref{FigOptimizationResults} B shows that the optimized networks for the three different initial networks commonly exhibit a characteristic topology, which has a densely interconnected core nodes and peripheral nodes dangling the core module. 
The heterogeneous network features were also observed in the basic network properties in the optimized networks (\TAB\ref{table:1}). Compared with the initial networks, the optimized networks tended to show larger averaged values of the shortest path length, $\left< \ell \right>$, betweenness centrality, $\left< b \right>$, and degree correlation, $r_{\rm assortative}$. The averaged values of the clustering coefficients, $\left< C \right>$, were smaller in the optimized networks. These results suggest that, through the optimization process, the network seems to enlarge its heterogeneity.

\subsection{Rich-club organization}\label{sec:RichClub}
This heterogeneous network topology has been reported as "rich-club" organization in a series of previous theoretical and experimental studies \cite{Zhou2004IEEEComm, Colizza2006NatPhys, GomezGardenes2010PLosOne, Heuvel2011JNS}. The prior literatures have characterized the organization by estimating normalized rich-club coefficients, $\Phi_{\rm norm} (k)$ described in \EQ\ref{eq:RichClubCoeffNormalize}. If the network has rich-club organization, $\Phi_{\rm norm} (k)$ should be more than $1$, and increase monotonically as $k$ increases.

\begin{figure}
\begin{center}
\includegraphics[width=15cm]{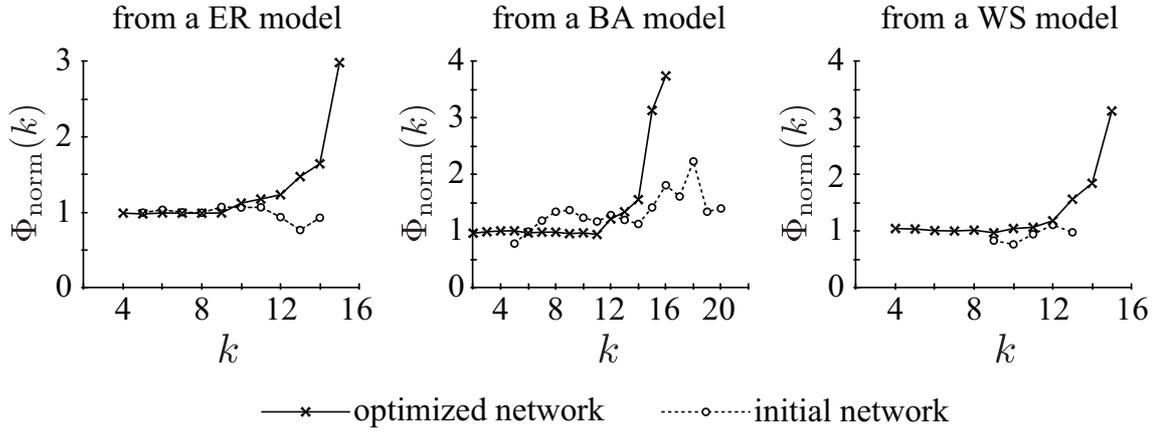}
\caption{
Difference in rich-club coefficients between the initial networks (circles and dashed lines) and optimized networks (multiple marks and solid lines).
While the normalized rich-club coefficients $\Phi_{\rm norm} (k)$ do not show monotonic increase in the initial networks, those in the optimized networks monotonically increase. These results suggest that the rewiring-based optimization changes the initial networks to networks with rich-club organization. To clarify the difference between before and after optimization, we adopted larger networks ($N = 100, \left<k\right> = 4$) than in \FIG\ref{FigOptimizationResults} ($N = 50, \left<k\right> = 4$).
}
\label{FigRichClub}
\end{center}
\end{figure}

\FIG\ref{FigRichClub} shows the comparison in $\Phi_{\rm norm} (k)$ between initial networks and optimized networks. To clarify the difference in the rich-club coefficients, we adopted as the initial networks larger networks than shown in \FIG\ref{FigOptimizationResults} B (\textit{i.e.,} $N = 100, \left<k\right> = 4$).
Before the optimization,  $\Phi_{\rm norm} (k)$ was not always larger than $1$ (\textit{e.g.,} ER and WS models), and did not show monotonic increase along $k$, which is consistent with a previous study \cite{Colizza2006NatPhys}.
In contrast, in the optimized networks, the rich-club coefficients were larger than $1$ in almost all the range of $k$, and monotonically increased as $k$ increased. These phenomena were observed commonly among the three different optimized networks that were derived from the three different initial networks.
In addition to the appearance of the networks in \FIG\ref{FigOptimizationResults} B, this estimation of $\Phi_{\rm norm} (k)$ supports the notion that the networks with rich-club organization has the minimum or a very small amount of synchronization cost due to phase difference among frequency-synchronized oscillators.

\subsection{Bimodal Degree Distribution}\label{sec:BimodalDegree}
As shown in \FIG\ref{FigOptimizationResults} B, the rich-club network topology consists of high-degree nodes cluster and low-degree peripheral nodes. Therefore, we hypothesized that the topology can be characterized by a bimodal degree distribution. 
To test the hypothesis, we estimated Wolfson's polarization index,  $\hat{P}$  \cite{Wolfson1997RIW}.
The Wolfson's index for degree distribution is defined as:
\begin{equation}
	\hat{P} = \frac{2 \left< k \right> }{m} \left( 2\left( \left<k\right>_2 - \left<k\right>_1 \right) - G \right),
\end{equation}
where $\left<k\right>$ is the mean of the degree, $k_i$, and $m$ denotes the median of the degree.
$\left<k\right>_1$ and $\left<k\right>_2$ are the mean values of $\{ k_i \bigm| k_i < m \}$ and of $\{ k_i \bigm| k_i \geq m \}$, respectively.
$G$ represents Gini inequality index, which is defined as $G = \frac{1}{2\left<k\right>} \sum_{i = 1}^{N} \sum_{j = 1}^{N} \bigm| k_i - k_j \bigm| $.
This Wolfson's polarization index shows the extent of the bimodality of the distribution. If the distribution is completely the same as a uniform distribution, the $\hat{P}$ is $0$. If the half of population has nothing and the other half shares everything, the $\hat{P}$ reaches a maximum, $0.25$.
In the present case, a larger $\hat{P}$ indicates that the network has a more bimodal and bipolarized degree distribution.

\begin{figure}
\begin{center}
\includegraphics[width=15cm]{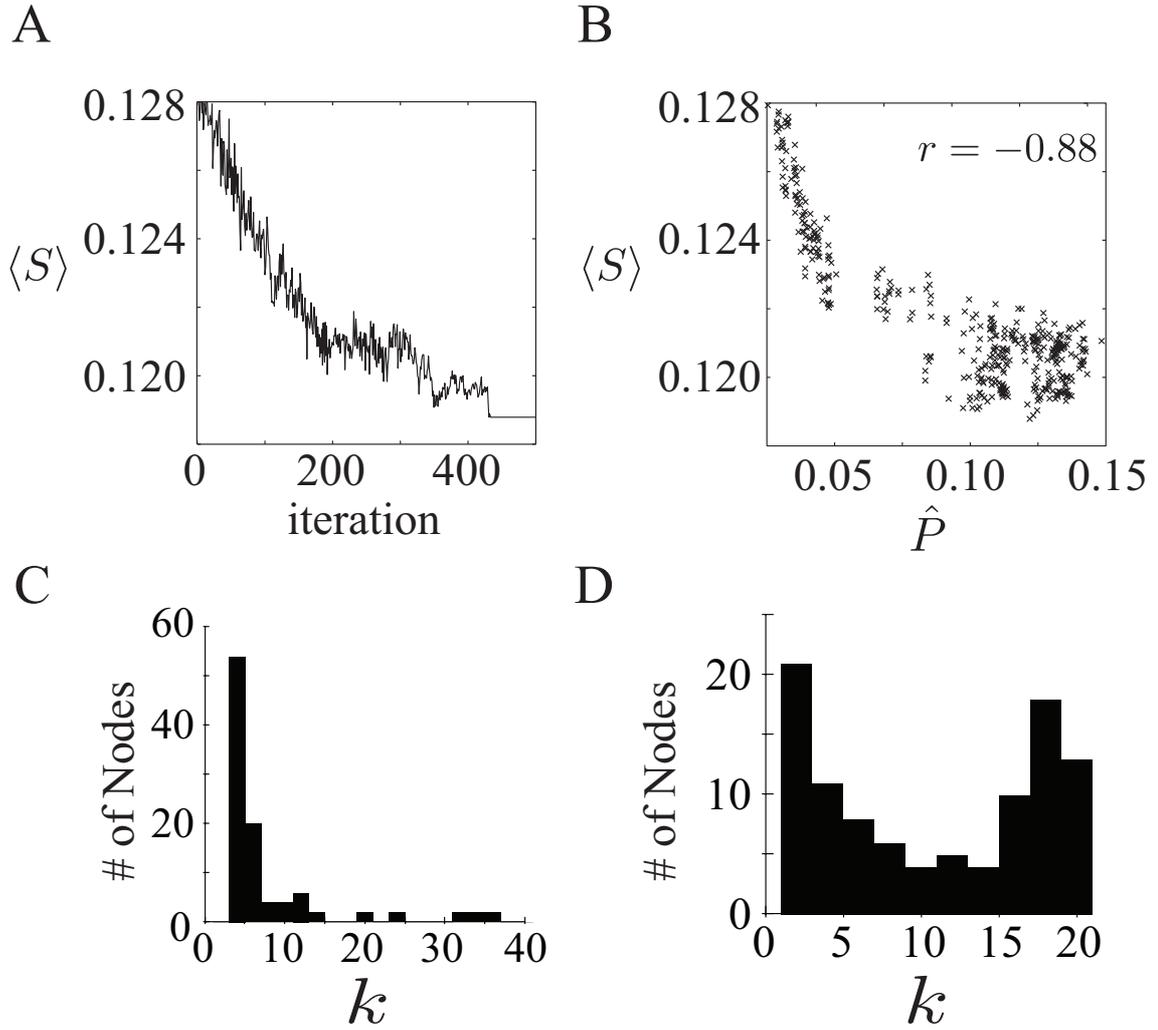}
\caption{
As the synchronization cost, $\left<S\right>$, decreases in the rewiring-based optimization (panel {\bf A}), 
the Wolfson's polarization index, $\hat{P}$, increases (panel {\bf B}).
This relation suggests that the rich-club network with a small amount of the synchronization cost can be characterized by bimodal degree distribution, which is quantified by Wolfson's polarization index.
Indeed, the degree distribution changed from a power-law distribution (panel {\bf C}) to a bimodal distribution (panel {\bf D}).
To clarify the difference between before and after optimization, we adopted larger networks ($N = 100, \left<k\right> = 4$) than in \FIG\ref{FigOptimizationResults} ($N = 50, \left<k\right> = 4$).
}
\label{FIgSyncVsPola}
\end{center}
\end{figure}

We estimated $\hat{P}$ during the rewiring-based optimization. 
Because $\hat{P}$ can be calculated more accurately for networks with more nodes, we used the BA model with $N = 100$ and $\left< k \right> = 4$ as an initial network for the optimization.
As a result, in the course of the above-mentioned optimization, $\left< S \right>$ decreased during the rewiring-based optimization (\FIG\ref{FIgSyncVsPola}A). 
Meanwhile, as $\left< S \right>$ decreases, $\hat{P}$ increases (\FIG\ref{FIgSyncVsPola}B).
Actually, the degree distribution changed from a power-law distribution (\FIG\ref{FIgSyncVsPola}C) to a bimodal distribution (\FIG\ref{FIgSyncVsPola}D).
This relation was also observed for different initial networks (e.g., ER model).
This correlation supports the hypothesis that rich-club network with small $\left< S \right>$ can be characterized by its bimodal degree distribution.

\subsection{Analytical Interpretation}\label{sec:AnalyticalExplanation}
\begin{figure}
\begin{center}
\includegraphics[width=15cm]{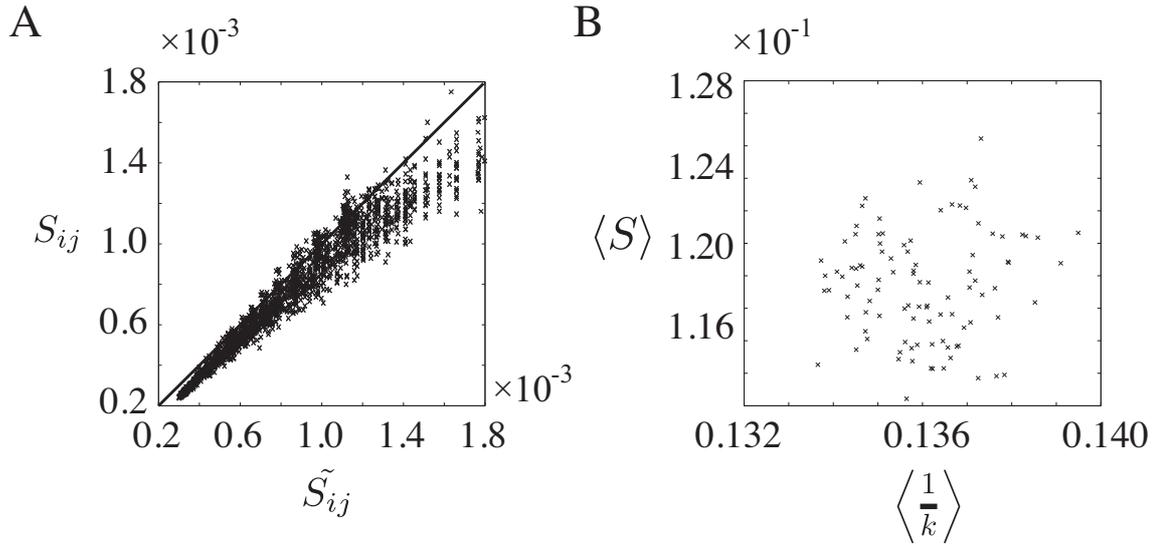}
\caption{
{\bf A} The analytical approximation of synchronization cost, $\tilde{ \left< S_{ij} \right>}$, is predictive of the real synchronization cost, $\left< S_{ij} \right>$, in the BA model with $N = 200, \left<N\right> = 10$. 
{\bf B}  Simple calculations using $\tilde{ \left< S_{ij} \right>}$ suggest a positive linear relationship between $\left< S \right>$ and $\left< \frac{1}{k} \right>$. However, there is not a strong correlation between them, which suggests a limitation of the approximation.
}
\label{FigSyncLoadApp}
\end{center}
\end{figure}

We finally provide an analytical interpretation on why the rich-club network has less $\left< S \right>$.
Using mean-field approximation, the Eq.~\eqref{eq:KuramotoEq} in the frequency-synchronized status can be described as $\Omega = \omega_i - \epsilon k_i \sin(\theta_i - \psi)$, 
where $\psi$ is defined in the Eq. ~\eqref{eq:OrderEq}. 
Therefore, if $ | \theta_i - \psi | $ is small enough, $ \theta_i - \psi = \frac{1}{\epsilon k_i}(\omega_i - \Omega) $, and $ (\theta_i - \theta_j)^2 $ is described as
\begin{equation}
	 (\theta_i - \theta_j)^2  = \frac{ (\omega_i - \Omega )^2 }{ {\epsilon}^2 {k_i}^2 } 
	 + \frac{ (\omega_j - \Omega )^2 }{ {\epsilon}^2 {k_j}^2 } 
	 - \frac{ 2 (\omega_i - \Omega )(\omega_j - \Omega ) }{ {\epsilon}^2 k_i k_j } ,
	 \label{eq:SyncLoad01}	 
\end{equation}
for a set of $ \{ \omega_i \} $.
 $\left< S_{ij} \right>$ is obtained as averaged $ (\theta_i - \theta_j)^2 $ across an enough number of sets of $ \{ \omega_i \} $. 
As in the above-described numerical estimation, we assume that $ \{ \omega_i \} $ is distributed according to the normal distribution with a mean of $\omega_0$ and a standard deviation of $\sigma$, and that the synchronized frequency is always $\Omega$ in every set of $ \{ \omega_i \} $.
Because we can also assume that $\omega_0$ is nearly equal to $\Omega$, $\left< (\omega_i - \Omega )^2 \right>$ is considered to be equal to $\sigma^2$, and $\left< (\omega_i - \Omega )(\omega_j - \Omega ) \right>$ is considered to be equal to zero.
Consequently, we obtain the approximation of $\left< S_{ij} \right>$ as follows:
\begin{equation}
	 \tilde{\left< S_{ij} \right>}   = \frac{\sigma^2}{ {\epsilon}^2} ( \frac{1}{{k_i}^2} + \frac{1}{{k_j}^2}). 
	  \label{eq:SyncLoad02}	
\end{equation}
This approximation was validated through comparison of $\tilde{\left< S_{ij} \right>}$ with the real $\left< S_{ij} \right>$, as shown in \FIG\ref{FigSyncLoadApp} A ($\sigma = 1, \epsilon = 0.3$). The two parameters had a large negative value of Pearson's correlation coefficient ($-0.88$).  

This expression of $ \tilde{\left< S_{ij} \right>}$ gives qualitative explanation on relationship between rich-club network topology and a small value of $\left<S\right>$:
To achieve a small amount of $\left<S\right>$, $ \tilde{\left< S_{ij} \right>}$ should be small.
When the node $i$ has a high degree, it makes more contribution to a smaller $ \tilde{\left< S_{ij} \right>}$ to connect with a the node $j$ with a high degree. It is also case when the node $i$ has a small degree.
As a result, for a small value of $ \tilde{\left< S_{ij} \right>}$, high degree nodes should connect with other high degree nodes, and low degree nodes should not have edges with other low degree nodes, but with high degree nodes. Consequently, high degree nodes tend to be gathered and create a densely-connected core module, and low degree nodes tend to connect with high degree nodes in the core module.
Overall, the optimized network with a small amount of $\left<S\right>$ is likely to be a rich-club network.

Note that it is difficult to further extend this approximation. If this approximation of $\left<S_{ij}\right>$ is accurate enough, a simple calculation of \EQ\ref{eq:SyncLoad02} leads us to the proportional relationship between $\left<S\right>$ and $\frac{1}{\left<k\right>} \left< \frac{1}{k} \right>$.
Given $\left<k\right>$ is a constant value as in the present study, $\left<S\right>$ should be proportional to $ \left< \frac{1}{k} \right>$. However, as shown in \FIG\ref{FigSyncLoadApp} B, we could not observe a linear relationship. 
This inaccurate approximation of $\left< S \right>$ may be caused by accumulation of the small difference between $\left<S_{ij}\right>$ and $\tilde{\left< S_{ij} \right>}$. This result suggests that we cannot extend this approximation to represent $\left<S\right>$ only by $\left< \frac{1}{k} \right>$.

\section{Discussion}
The present study introduced synchronization cost based on phase difference among frequency-synchronized oscillators.
Using the rewiring-based optimization \cite{Donetti05PRL, Gorochowski2010PRE}, we showed that the synchronization cost is minimized in a rich-club network topology.
Furthermore, we demonstrated that the network topology can be characterized by the bimodality of its degree distribution. 
Finally, we provided analytical explanation on why the rich-club network topology is associated with a small amount of synchronization cost.
%

The concept of synchronization cost is not a novel idea of the present study. As described in \SEC\ref{Sec:Introduction}, a line of previous studies have investigated a different type of synchronization cost, which is based on coupling strength \cite{Motter2005PRE, Nishikawa2006PhyD, Brede2008EPJB}. Whereas the present synchronization cost due to phase difference can be regarded as dynamic cost per unit time, the cost based on coupling strength can be considered as static cost that is related to building and maintaining of network infrastructures. Interestingly, the optimal network topology with the least cost depends on which of the two types of synchronization cost we adopt. The optimal networks for the synchronization cost based on coupling strength often show more homogeneous properties \cite{Motter2005PRE} than those for the other synchronization cost. The homogeneity of networks is desired to enhance synchronizability \cite{Nishikawa2003PRL, Donetti05PRL}. Therefore, it may be necessary to investigate what network structures balance these two types of synchronization cost. 

The present synchronization cost in the present study can be another concept of load assigned to edge in a complex network.
Previous studies used edge-betweenness as edge load \cite{Holme2002PRE, Guimera2003PRE}, 
which is useful in various situations from human interaction \cite{Guimera2003PRE} to data transmission in computer networks \cite{Holme2002PRE}. 
However, because the edge betweenness does not consider synchrony in networks, its properties have evident difference from those of the synchronization cost.
For example, as shown in \SEC\ref{sec:AnalyticalExplanation}, the synchronization cost is lower between high degree nodes, and higher between low degree nodes. 
In contrast, the edge betweenness tends to be higher in edges bridging high degree nodes and be lower in edges bridging low degree nodes.
These distinct properties suggest the possibility that the synchronization cost can be another concept of edge load.

The synchronization cost in the present study, $S_{ij}$, has a mathematical expression similar to that for local synchronizability, $r_{\rm local}$ \cite{Gomez-Gardenes2007PRLPRE}. However, the two parameters focus on different phases of synchronization in complex networks. 
The local synchronizability enables us to quantify the local construction of the synchronization pattern. Therefore, it is useful to investigate properties of networks that are not yet fully synchronized. 
In contrast, the synchronization cost in the present study can be only estimated in fully-synchronized networks. Therefore, in the present study, we used a relatively large coupling strength, and achieved full synchronization throughout the optimizations. 
As a result, in the present study, the local synchronizability was always saturated.

Although the present study did not adopt models specific to any real networks, the findings may help understanding  large-scale brain networks. Recently, a few of recent studies have reported the existence of the rich-club organization in the large-scale brain networks. A previous empirical study has demonstrated the existence of the rich-club organization in the large-scale human brain networks \cite{Heuvel2011JNS}. Another study has investigated the anatomical connectivity in the cerebral cortex of cats, and has showed that rich-club organization controls the dynamic transition of synchronization in the brain \cite{GomezGardenes2010PLosOne}. A recent review has suggested that the organization is a cost-effective network topology for the brain networks, which are required to be adapted to various cognitive functions \cite{Bullmore2012NatRevNeuro}. 
In addition to the context of cost-effectiveness, the rich-club network topology is robust to random attack \cite{Paul2006PhysicaA}. This previous study analytically and numerically demonstrated that networks robust to random attack have similar structures observed in the present study. The robust networks highly-interconnected hub modules and peripheral nodes (\textit{leaf} nodes) that have a single edge. This network topology has bimodal degree distribution and shows rich-club organization. 
Considering these prior literatures, it is suggested that the rich-club organization is beneficial for the large-scale brain networks to efficiently and robustly maintain its wide range of functions based on synchronization.

\begin{acknowledgements} 
The author acknowledges the support from the Japan Society for the Promotion of Science (JSPS) Research Fellowship for Young Scientists (222882).
\end{acknowledgements}


\clearpage
\newpage

\pagestyle{empty}

\clearpage

\clearpage

\clearpage

\clearpage

\begin{table}
\caption{Basic topological properties of the optimized networks.
Despite different initial networks, the optimized networks had similar network topological properties. The values for the initial networks represent averaged values across ten estimations, whereas the values for the optimized networks show the mean $\pm$ s.d. across the ten estimations.
}
\begin{center}

\begin{tabular}{ccccccccc} \toprule 
  
& \multicolumn{2}{c}{$\left< \ell \right>$} & \multicolumn{2}{c}{$\left< C \right>$} & \multicolumn{2}{c}{$\left< b \right>$} & \multicolumn{2}{c}{$ r_{\rm assortative} $}  \\ 
& initial & optimized & initial & optimized & initial & optimized & initial & optimized \\ \midrule
From ER models & 1.9& 3.5 $\pm$ 0.025 & 0.21 & 0.13 $\pm$ 0.010 & 44 & 122.2 $\pm$ 3.7 & 0.034 & 0.25 $\pm$ 0.12 \\
From BA models & 1.8 & 3.3 $\pm$ 0.021 & 0.33 & 0.15 $\pm$ 0.012 & 48 & 113 $\pm$ 2.8 & -0.15 & 0.28 $\pm$ 0.021 \\
From WS models & 2.3 & 3.1 $\pm$  0.017 & 0.62 & 0.16 $\pm$ 0.010  & 63 & 105 $\pm$ 2.4 & 0.047  & 0.25 $\pm$ 0.011 \\ \bottomrule
 
 \end{tabular}

\label{table:1}
\end{center}

\end{table}

\clearpage

\appendix

\section{Definition of Synchronization Cost}\label{Appendix: SyncCostInPowerGrid}
In this section, we explain why power loss consumed in the electric line between power plants can be represented by square of difference in phase of voltage between the two power plants.

\begin{figure}
\begin{center}
\includegraphics[width=15cm]{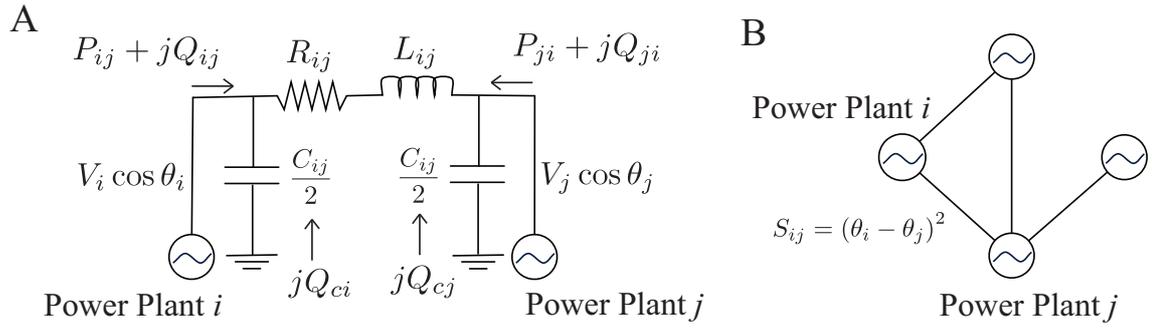}
\caption{
Panel {\bf A} shows a typical power line between power plants $i$ and $j$. 
$V_i \cos{\theta_i}$ and $V_j \cos{\theta_j}$ indicate the voltage of the output from the power plants. 
$R_{ij}$, $L_{ij}$, and $C_{ij}$ indicate resistance, inductance, and conductance between the power plants. 
As show in panel {\bf B}, $S$ is defined in every power line based on the phase difference of the voltages between the connecting power plants. 
}
\label{FigPowerLine}
\end{center}
\end{figure}

In the following model, as in previous studies \cite{Fioriti2009CIIS, Dorfler2010ACC}, we do not consider the effect of the length of the power line on the power loss.
To estimate the power loss in a typical power line shown in \FIG\ref{FigPowerLine}, we estimate active power flow ($P_{ij}$ and $P_{ji}$), reactive power flow ($Q_{ij}$ and $Q_{ji}$), and delayed reactive power flow ($Q_{ci}$ and $Q_{cj}$) as follows \cite{Kundur1993Elgerd1982Book}:
\begin{gather}
	P_{ij} = \frac{ V_i V_j \sin(\theta_i - \theta_j) }{ {Z_{ij}}^2 / f_0 L_{ij} } 
	+ \frac{ {V_i}^2 - V_i V_j \cos(\theta_i - \theta_j) }{ {Z_{ij}}^2 /R_{ij} },  \\
	Q_{ij} = - \frac{ V_i V_j \sin(\theta_i - \theta_j) }{ {Z_{ij}}^2 / R_{ij} } 
	+ \frac{ {V_i}^2 - V_i V_j \cos(\theta_i - \theta_j) }{ {Z_{ij}}^2 / f_0 L_{ij}},   \\
	Q_{ci} = \frac{ f_0 C_{ij}}{2} {V_i}^2, 
\end{gather}
where $f_0$ represents synchronized angular frequency of alternating voltage, and ${Z_{ij}}^2 = {R_{ij}}^2 + ( f_0 L_{ij} )^2 $. $P_{ji}$, $Q_{ji}$, and $Q_{cj}$ are obtained by exchanging $i$ and $j$.
Using these power flows, the active power loss due to resistance, $P_{\rm loss}^{ij}$, is calculated as $P_{ij} + P_{ji}$, whereas the reactive power loss due to inductance, $Q_{\rm loss}^{ij}$, is estimated as $Q_{ij} + Q_{ji} + Q_{ci} + Q_{cj}$ as follows:
\begin{gather}
	P_{\rm loss}^{ij} =  \frac{ R_{ij} }{{Z_{ij}}^2 }
			 \left(- 2 V_i V_j \cos(\theta_i - \theta_j) + {V_i}^2 + {V_j}^2  \right),  \\
	Q_{\rm loss}^{ij} = \frac{ {Z_{ij}}^2 }{f_0 L_{ij} }
	 		\left(- 2 V_i V_j \cos(\theta_i - \theta_j) + {V_i}^2 + {V_j}^2 \right) 
	 		+ \frac{f_0 C_{ij}}{2} ({V_i}^2 + {V_{j}}^2).
\end{gather}
The total power loss is estimated as a combination of the active power loss and the reactive power loss \cite{Kundur1993Elgerd1982Book}. 
By using a second-order Taylor expansion, we regard the total power loss, $P_{\rm loss}^{ij} + Q_{\rm loss}^{ij}$, as $a_0 + a_1 \left( \theta_i - \theta_j \right)^2$, where $a_0$ and $a_1$ are constants ($a_1 > 0$).
 Therefore, we define the synchronization cost, $S_{ij}$, for a power line between power plants $i$ and $j$ as 
\begin{equation}
	S_{ij} = \left( \theta_i - \theta_j \right)^2.
\end{equation}

\section{Power Grid as Kuramoto Model}\label{Appendix: PowerGridAsKuramotoModel}
In this section, we explain that, under several assumptions, we can approximate power grids by the first-order Kuramoto model of nonuniform oscillators.

As previous studies \cite{Buzna2009PRE, Fioriti2009CIIS, Dorfler2010ACC}, we model a power grid as follows: 
The structure of the power grid with $N$ power plants is represented as an unweighted and undirected adjacency matrix $A$, 
where a node represents a power plant and an edge a power line. 
$A_{ij}$ is $1$ when power plants $i$ and $j$ have a power line between them, and $A_{ij}$ is $0$ when they do not. 
According to the previous studies \cite{Buzna2009PRE, Fioriti2009CIIS, Dorfler2010ACC}, the phase of the output voltage of the power plant $i$, $\theta_i$, is described as 
\begin{equation}
	 \dot{\theta_i} = \frac{f_i}{D_i} - \sum_{j}{ \frac{W_{ij}}{D_i} A_{ij} \sin(\theta_i - \theta_j)},
\end{equation}
where 
$D_i$ denotes a damping constant, 
$W_{ij}$ is an amount of power transfer between power plants $i$ and $j$, 
and $f_i$ represents the natural frequency of the output voltage from the power plant $i$.
To reduce computational cost for the following rewiring-based optimization, we assume that $W_{ij}/D_i = \epsilon$ for any power line. 
Because ${f_i}/{D_i}$ is specific to power plant $i$, we replace the value with $\omega_i$. 
Consequently, the voltage phase of the power plants can be expressed in the Kuramoto model as
\begin{equation}
	 \dot{\theta_i} = {\omega_i} - \epsilon \sum_{j}{A_{ij} \sin(\theta_i - \theta_j)}.
\end{equation}

\clearpage

\end{document}